\documentstyle[12pt]{article}

\font\twlgot =eufm10 scaled \magstep1
\font\egtgot =eufm8
\font\sevgot =eufm7

\font\twlmsb =msbm10 scaled \magstep1
\font\egtmsb =msbm8
\font\sevmsb =msbm7

\newfam\gotfam

\textfont\gotfam\twlgot
\scriptfont\gotfam\egtgot
\scriptscriptfont\gotfam\sevgot

\newfam\msbfam
\textfont\msbfam\twlmsb
\scriptfont\msbfam\egtmsb
\scriptscriptfont\msbfam\sevmsb

\def\pBbb{\relax\ifmmode\expandafter\Bb\else\typeout{You cann't use
Bbb in text mode}\fi}
\def\Bb #1{{\fam\msbfam\relax#1}}

\def\thebibliography#1{\bigskip\section*{\large
\bf References\\}\list
  {[\arabic{enumi}]}{\settowidth\labelwidth{#1}\leftmargin\labelwidth
    \advance\leftmargin\labelsep
    \usecounter{enumi}}
    \def\newblock{\hskip .11em plus .33em minus .07em}
    \sloppy\clubpenalty4000\widowpenalty4000
    \sfcode`\.=1000\relax}

\def\op#1{\mathop{{\it\fam0} #1}\limits}

\newcommand{\beq}{\begin{equation}}
\newcommand{\eeq}{\end{equation}}
\newcommand{\ben}{\begin{eqnarray}}
\newcommand{\een}{\end{eqnarray}}
\newcommand{\be}{\begin{eqnarray*}}
\newcommand{\ee}{\end{eqnarray*}}
\newcommand{\bea}{\begin{eqalph}}
\newcommand{\eea}{\end{eqalph}}

\newcommand{\cL}{{\cal L}}

\newcommand{\cE}{{\cal E}}

\newcommand{\bL}{{\bf L}}

\newcommand{\dl}{\delta}
\newcommand{\la}{\lambda}

\newcommand{\vf}{\varphi}

\newcommand{\om}{\omega}

\newcommand{\m}{\mu}

\newcommand{\th}{\theta}

\newcommand{\si}{\sigma}

\newcommand{\w}{\wedge}

\newcommand{\dr}{\partial}

\newcommand{\ap}{\approx}

\newenvironment{eqalph}{\stepcounter{equation}
\setcounter{equationa}{\value{equation}}
\setcounter{equation}{0}

\begin{eqnarray}}{\end{eqnarray}\setcounter{equation}{\value{equationa}}}

\newcommand{\mar}[1]{}

\begin{document}
\hbox{}

{\parindent=0pt

{\large \bf Noether conservation laws issue from the gauge invariance of
an Euler--Lagrange operator, but not a Lagrangian}
\bigskip 

{\bf G.Sardanashvily}

\medskip

\begin{small}

Department of Theoretical Physics, Moscow State University, 117234
Moscow, Russia

E-mail: sard@grav.phys.msu.su

URL: http://webcenter.ru/$\sim$sardan/
\bigskip

{\bf Abstract.}
As is well known, there are different Lagrangians which lead to the
same Euler--Lagrange operator. The gauge invariance of a Lagrangian 
guarantees that of
the corresponding Euler--Lagrange operator, but not {\it vice versa}.
We show that the gauge invariance of an Euler--Lagrange operator, but not
a Lagrangian results in Noether conservation laws.

\end{small}
}

\bigskip
\bigskip

Let us consider a first order field theory on a smooth fibre bundle $Y\to
X$ coordinated by $(x^\la,y^i)$. Its configuration space is the first
order jet manifold $J^1Y$ of sections of $Y\to X$ equipped with the
adapted coordinates $(x^\la,y^i,y^i_\m)$ (see, e.g., \cite{book,epr}).
We will use the notation $\om=d^nx$, $n=\dim X$, and
$\om_\la=\dr_\la\rfloor \om$.

A first order Lagrangian
is a density
\mar{cc201}\beq
L=\cL(x^\la,y^i,y^i_\la)\om: J^1Y\to\op\w^nT^*X \label{cc201}
\eeq
on $J^1Y$. Its variation $\dl L$ is the second order Euler--Lagrange operator
\mar{305}\ben
&& \cE_L: J^2Y\to T^*Y\w(\op\w^nT^*X), \nonumber \\
&& \cE_L= (\dr_i\cL- d_\la\dr^\la_i)\cL \th^i\w\om, \label{305}
\een
where $\th^i=dy^i-y^i_\la dx^\la$ are contact forms and $d_\la=\dr_\la
+y^i_\la\dr_i +y^i_{\la\m}\dr_i^\m$ are the total 
derivative. Its kernel Ker$\,\cE_L\subset J^2Y$ defines the Euler--Lagrange 
equations. There are different Lagrangians whose variations provide the
same Euler--Lagrange operator. They make up an affine space
modelled over the vector space of variationally trivial Lagrangians
$L_0$, i.e., $\dl L_0=0$. One can show that a first order
Lagrangian $L_0$ is variationally trivial iff 
\mar{mos11}\beq
L_0=h_0(\vf), \label{mos11}
\eeq
where
$\vf$ is a closed $n$-form on $Y$ and 
\be
h_0(dx^\la)=dx^\la, \qquad h_0(\th^i)=0
\ee
is the horizontal operator acting on semibasic exterior forms on
$J^1Y\to Y$. 

Gauge transformations are defined as bundle automorphisms of
a fibre bundle $Y\to X$. 
Any projectable
vector field 
\mar{v1}\beq
u=u^\la(x^\m)\dr_\la + u^i(x^\m,y^j)\dr_i \label{v1}
\eeq
on $Y\to X$ is the infinitesimal generator of a 
local one-parameter group $G_u$ of gauge transformations of $Y\to X$, and
{\it vice versa}. Its jet prolongation onto $J^1Y$ reads
\mar{1.21}\beq
J^1u =u^\la\dr_\la + u^i\dr_i + (d_\la u^i
- y_\m^i\dr_\la u^\m)\dr_i^\la. \label{1.21}
\eeq
A Lagrangian $L$ (\ref{cc201}) is invariant under a gauge
group $G_u$ iff its Lie derivative 
\mar{04}\beq
\bL_{J^1u}L=[\dr_\la (u^\la\cL) +(u^i\dr_i 
+(d_\la u^i -y^i_\m\dr_\la u^\m)\dr^\la_i)\cL]\om \label{04}
\eeq
along $J^1u$ vanishes.

The first variational formula of the calculus of variations provides
the canonical decomposition 
\mar{bC30'}\beq
\bL_{J^1u}L= u_V\rfloor \cE_L + h_0 d(u\rfloor H_L) \label{bC30'}
\eeq
of the Lie derivative (\ref{04}),
where: $u_V=(u\rfloor\th^i)\dr_i$, $\cE_L$ is the Euler--Lagrange
operator (\ref{305}) and $H_L$ is some Lepagean equivalent
of $L$, e.g., the Poincar\'e--Cartan form 
\be
 H_L=L +\dr^\la_i\cL\th^i\w\om_\la=\dr^\la_i\cL dy^i\w\om_\la +
(\cL- y^i_\la\dr^\la_i\cL)\om.
\ee
For instance, if $L=L_0$ is a variationally trivial Lagrangian, the
first variational formula (\ref{bC30'}) gives the equality
\mar{v4}\beq
\bL_{J^1u}L_0=h_0 d(u\rfloor H_{L_0}). \label{v4}
\eeq

If the Lie derivative (\ref{04}) vanishes, one obtains the weak
conservation law
\mar{v2}\beq
0\ap h_0 d(u\rfloor H_L)=d_\la[\pi^\la_i(u^\m y^i_\m -u^i)
-u^\la\cL]\om \label{v2} 
\eeq
on Ker$\,\cE_L$.

The Euler-Lagrange operator $\cE_L$ (\ref{305})
is invariant under a one-parameter gauge group
$G_u$ iff its Lie derivative  $\bL_{J^2u}\cE_L$ along the
jet prolongation $J^2u$ of $u$ onto the second order jet manifold $J^2Y$ 
vanishes. There is the relation
\mar{v3}\beq
\bL_{J^2u}\cE_L=\dl(\bL_{J^1u}L)=\cE_{\bL_{J^1u}L}, \label{v3}
\eeq 
i.e., the Lie derivative $\bL_{J^2u}\cE_L$ is the Euler--Lagrange
operator associated to the Lagrangian $\bL_{J^1u}L$ \cite{giac90,book}.
It follows that 
$\bL_{J^2u}\cE_L=0$ iff the Lie derivative $\bL_{J^1u}L$ is a
variationally trivial Lagrangian, i.e., is given by the expression
(\ref{mos11}). 
Substituting this expression into the first variational formula
(\ref{bC30'}), we obtain the weak equality
\mar{v5}\beq
h_0(\vf)\ap h_0 d(u\rfloor H_L)\label{v5}
\eeq
on Ker$\,\cE_L$. If $\vf=d\si$ is an exact form on $Y$, this equality is
brought into the weak conservation law
\mar{v6}\beq
0\ap h_0 d(u\rfloor H_L-\si). \label{v6}
\eeq

Let $L'=L+L_0$ be another Lagrangian whose variation is the
Euler--Lagrange operator $\cE_L$. Due to the equality (\ref{v4}), we
come to the same conservation law (\ref{v6}).

\end{document}